\documentclass[useAMS,usenatbib,SFB@referee]{mn2e}

\def\lesssim{\mathrel{\hbox{\rlap{\hbox{\lower4pt\hbox{$\sim$}}}\hbox{$<$}}}}
\def\gtrsim{\mathrel{\hbox{\rlap{\hbox{\lower4pt\hbox{$\sim$}}}\hbox{$>$}}}}
\def\geqq{\mathrel{\hbox{\rlap{\hbox{\lower4pt\hbox{$=$}}}\hbox{$>$}}}}

\usepackage[dvips]{graphicx}
\input epsf

\title[Erratum to Star formation triggered by SN explosions]{Erratum to Star formation triggered by SN explosions: an application to the stellar association of $\beta$Pictoris}

\author[C. Melioli et al.]{C. Melioli$^{1,2}$\thanks{cmelioli@astro.iag.usp.br}, E. M. de Gouveia Dal Pino$^{1}$\thanks{dalpino@astro.iag.usp.br},
M. R. M. Le\~ao$^{1}$\thanks{mrmleao@astro.iag.usp.br}, R. de la Reza$^{3}$\thanks{delareza@on.br}, A. Raga$^{4}$\thanks{raga@nuclecu.unam.mx}\\
$^{1}$ Universidade de S\~ao Paulo, IAG, Rua do Mat\~ao 1226,Cidade Universit\'aria, S\~ao Paulo 05508-900, Brazil\\
$^{2}$ University of Bologna, Italy\\
$^{3}$ Observat\'orio Nacional, Rua General Jos\'e Cristino 77, S\~ao Cristov\~ao, 20921-400 Rio de Janeiro, Brazil\\
$^{4}$ Instituto de Ciencias Nucleares, Universidad Nacional
Aut\'onoma de M\'exico, Ap.P. 70543, 04510 DF, Mexico}

\begin{document}

\date{Accepted ??? ???. Received ??? ???; in original form ??? ??? ???}

\pagerange{\pageref{firstpage}--\pageref{lastpage}}
\pubyear{2006}

\maketitle

\label{firstpage}

\begin{abstract}
{This is an erratum to the article entitled "Star formation
triggered by SN explosions: an application to the stellar
association of $\beta$ Pictoris" by C. Melioli, E. M. de Gouveia Dal
Pino, R. de la Reza and A. Raga which was published in this Journal
(MNRAS, 373, 811-818, 2006) with the following original abstract:

In the present study, considering the physical conditions that are
relevant interactions between supernova remnants (SNRs) and dense
molecular clouds for triggering star formation we have built a
diagram of  SNR radius versus cloud density in which the
constraints above delineate a shaded zone where star formation is
allowed. We have also performed fully 3-D radiatively cooling
numerical simulations of the impact between SNRs and  clouds under
different initial conditions in order to follow the initial steps of
these interactions. We determine the conditions that may lead either
to cloud collapse and star formation or to complete cloud
destruction and find that the numerical results are consistent with
those of the SNR-cloud density diagram. Finally, we have applied the
results above to the $\beta-$ Pictoris stellar association which is
composed of low mass Post-T Tauri stars with an age of 11 Myr. It
has been recently suggested  that its formation could have been
triggered by the shock wave produced by a SN explosion localized at
a distance of about 62 pc that may have occurred either in the Lower
Centaurus Crux (LCC) or in the Upper Centaurus Lupus (UCL) which are
both nearby older subgroups of that association (Ortega and
co-workers).Using the results of the analysis above we have shown
that the suggested origin for the young association at the proposed
distance is plausible only for a very restricted range of initial
conditions for the parent molecular cloud, i.e., a cloud with a
radius of the order of 10 pc, a density of the order of 10$-$20
cm$^{-3}$, and a temperature of the order of 10$-$100 K }.
\end{abstract}


\section{The Erratum}

The original manuscript should be modified as follows:

\subsection{Alterations in Section 2.2}
In Section 2.2, the equation (12) that describes the curvature
effects on the shock velocity of the supernova remnant-cloud
interaction:

\begin{equation}
{\hat{v}}_{cs} = v_{snr} \ \left({{n_{sh}} \over {n_c}}\right)^{0.5}\ {1 \over t_{c,snr}} \ \int^{t_{c,snr}}_0{cos \gamma(t) \ dt}
\end{equation}

\noindent where \[   \cos\gamma(t)\;= \frac{d^2
-R_{snr}^2(t)-r_c^2}{2 R_{snr}^2(t) \; r_c} \; ,\]

\noindent should be more precisely integrated  up to:
 \[t_{c,snr}=\frac{r_c +
R_{snr}-\sqrt{r_c^2+R_{snr}^2}}{v_{snr}}\; .\]

\noindent This equation has  an exact solution that should replace
the approximate one given in the original manuscript by Figure 2.
With the integration limits above the exact solution of Eq. (12)
 is given by:

\[\hat{v}_{cs} = \left( \frac{n_{sh}}{n_{c}}\right)^{1/2} \; v_{snr} \;
 I\]

\noindent where

\begin{eqnarray}
I = A \left[ 4.8 R_{snr}^2 + 10.6 R_{snr} r_c - 0.5 r_c^2 \right. \nonumber
\end{eqnarray}

\begin{eqnarray}
\left. + (R_{snr} + 0.5 r_c )
 \sqrt{R_{snr}^2 + r_c^2} - B +  C \right] \nonumber
\end{eqnarray}

\noindent and

\[A = \frac{1}{r_c (R_{snr} + r_c - \sqrt{R_{snr}^2 + r_c^2})}\]
\[B =  R_{snr}\, (1.25 R_{snr} + 2.5 r_c)\log{R_{snr}}\]
\[C = R_{snr}\,(0.5 R_{snr} + r_c) \log{\left[R_{snr}^{3/2} D \right]}\]
\[D = \left(1.86 \times 10^{-5} R_{snr} + 9.3\times 10^{-6} r_c \right.\]
\[\left. - \; 9.3\times 10^{-6} \sqrt{R_{snr}^2 + r_c^2}\right)\]

The solution above to $\hat{v}_{cs}$ will produce slight changes on
the multiplying factors that appear in the equations (13) to (26) of
the original manuscript as indicated below:

\begin{equation}
t_{cc,A}\sim 5 \times 10^5 \; \frac{n_{c,10}^{0.5}\; r_{c,10} \; R_{snr,50}^{1.5}}{I_5 \; E_{51}^{0.5}} \ \ \ {\rm yr}
\end{equation}

\begin{equation}
t_{cc,R} \sim 5 \times 10^5 \; \frac{ r_{c,10} \;R_{snr,50}^{2.5}\; n_{c,10}^{0.5}\; n^{0.41}}{I_5\; f_{10}^{0.5}\; E_{51}^{0.8}} \ \ \ {\rm yr}
\end{equation}

\begin{equation}
M_A \approx 20\;\frac{E_{51}^{0.5}\;I_5}{T_{c,100}^{0.5}\; R_{snr,50}^{1.5}\; n_{c,10}^{0.5}}
\end{equation}

\begin{equation}
M_R \approx 22\;\frac{f_{10}^{0.5} \;E_{51}^{0.8}\;I_5}{n_{c,10}^{0.5}\; T_{c,100}^{0.5}\;R_{snr,50}^{2.5}\; n^{0.41}}
\end{equation}

\begin{equation}
n_{c,sh,A} \sim \frac{4000}{R_{snr,50}^3}\; \frac{E_{51}\;I_5^2}{T_{c,100}} \ \ \ {\rm cm^{-3}}
\end{equation}

\begin{equation}
n_{c,sh,R} \sim \frac{4800}{R_{snr,50}^5}\; \frac{E_{51}^{1.6}\;I_5^2\;f_{10}}{T_{c,100}\; n^{0.82}} \ \ \ {\rm  cm^{-3}}
\end{equation}

\begin{equation}
m_{J,A}\sim 1500 \frac{T_{c,100}^2\;R_{snr,50}^{1.5}}{I_5\;E_{51}^{0.5}} \;\; M_{\odot}
\end{equation}

\begin{equation}
m_{J,R}\sim 1400 \frac{T_{c,100}^2\;R_{snr,50}^{2.5}\;n^{0.41}}{I_5\;f_{10}^{0.5}\;E_{51}^{0.8}}\;\;M_{\odot}
\end{equation}

\begin{equation}
    r_{c,A}\geq 10.4\;\frac{T_{c,100}^{2/3}\;R_{snr,50}^{0.5}}{I_5^{1/3}\; n_{c,10}^{1/3}\;E_{51}^{0.17}}\ \ \ {\rm pc}
\end{equation}

\begin{equation}
    r_{c,R}\geq 10.2\; \frac{T_{c,100}^{2/3}\;_{snr,50}^{0.83}\; n^{0.14}}{I_5^{1/3} \;f_{10}^{0.17}\; E_{51}^{0.27}\; n_{c,10}^{0.3} \; r_{c,10}^{2.4}}\ \ \ {\rm pc}
\end{equation}

\subsection{Alterations in Section 3.2}

\ In Section 3.2, $t_{un} \le 3 t_{cc}$, should be replaced by
$t_{un} \le 5 t_{cc}$, since (as remarked in the original
manuscript) $5 t_{cc}$ is a more representative average value of the
time scale at which the density of the shocked material of the cloud
drops by a factor of two after the impact according to radiative
cooling numerical simulations (Melioli et al. 2005).

Also, the exponent of $r_c$ in Eq. (27)  should be replaced by
$7/3$, so that the correct equation reads:

\begin{equation}
M \le 19 \left(\frac{n_{c,10}}{T_{c,100}}\right)^{1.16} r_{c,10}^{7/3}
\end{equation}

\noindent This implies  modifications also in the $r_c$ exponents of Eqs. (28) and (29), as described below, respectively by:

\begin{equation}
R_{snr,A} \ge 52 \ {{E_{51}^{0.33} \ T_{c,100}^{0.44} \ I_5} \over{n_{c,10} \ r_{c,10}^{1.56}}} \ \ \ \ {\rm pc}
\end{equation}

\noindent and

\begin{equation}
R_{snr,R} \ge 53 \ {{E_{51}^{0.33} \ f_{10}^{0.2} \ T_{c,100}^{0.26}\ I_5^{0.4}} \over {n_{c,10}^{0.7} \ n^{0.17} \ r_{c,10}^{0.93}}} \ \ \ \ {\rm pc}
\end{equation}

\subsection{Alterations in Section 3.3}

In Section 3.3, there is a typo in eq. (31). The 9/16 factor should
be replaced by 25/8, i.e,:

\begin{equation}
t_{st} \simeq {25 \over 8} {{\mu  m_H} \over {n_c \Lambda}}v_{cs}^2\ \ \ {\rm s.}
\end{equation}

\noindent Considering this  correction and the one due to $ \hat{v}_{cs}$ in Eq. (12), the multiplying factor 75 in  eq. (32) should be replaced by 160, or:

\begin{equation}
R_{snr,A} \le 160 {{E_{51}^{0.33} I_5^{0.66}} \over {(r_{c,10}\Lambda_{27})^{2/9} n_c^{0.5}}} \ \ \ {\rm pc}
\end{equation}

The modifications above, particularly those in Eqs. (27-28) and
Eqs.(31-32) result slight modifications in the diagrams of Figure 3
which should be replaced by the  Figure below.

\begin{figure}
    \centering
        \includegraphics[width=6cm]{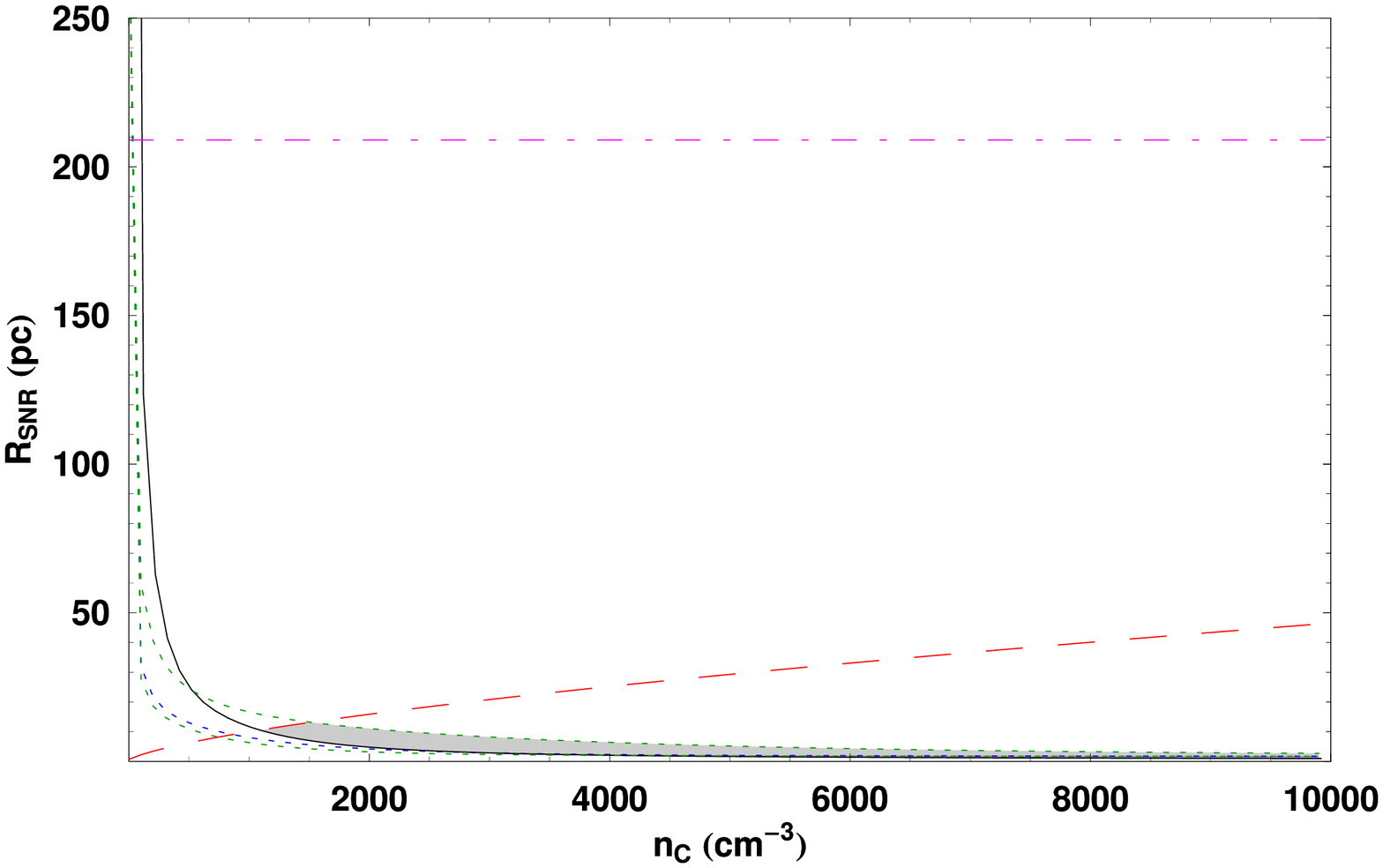}
        \includegraphics[width=6cm]{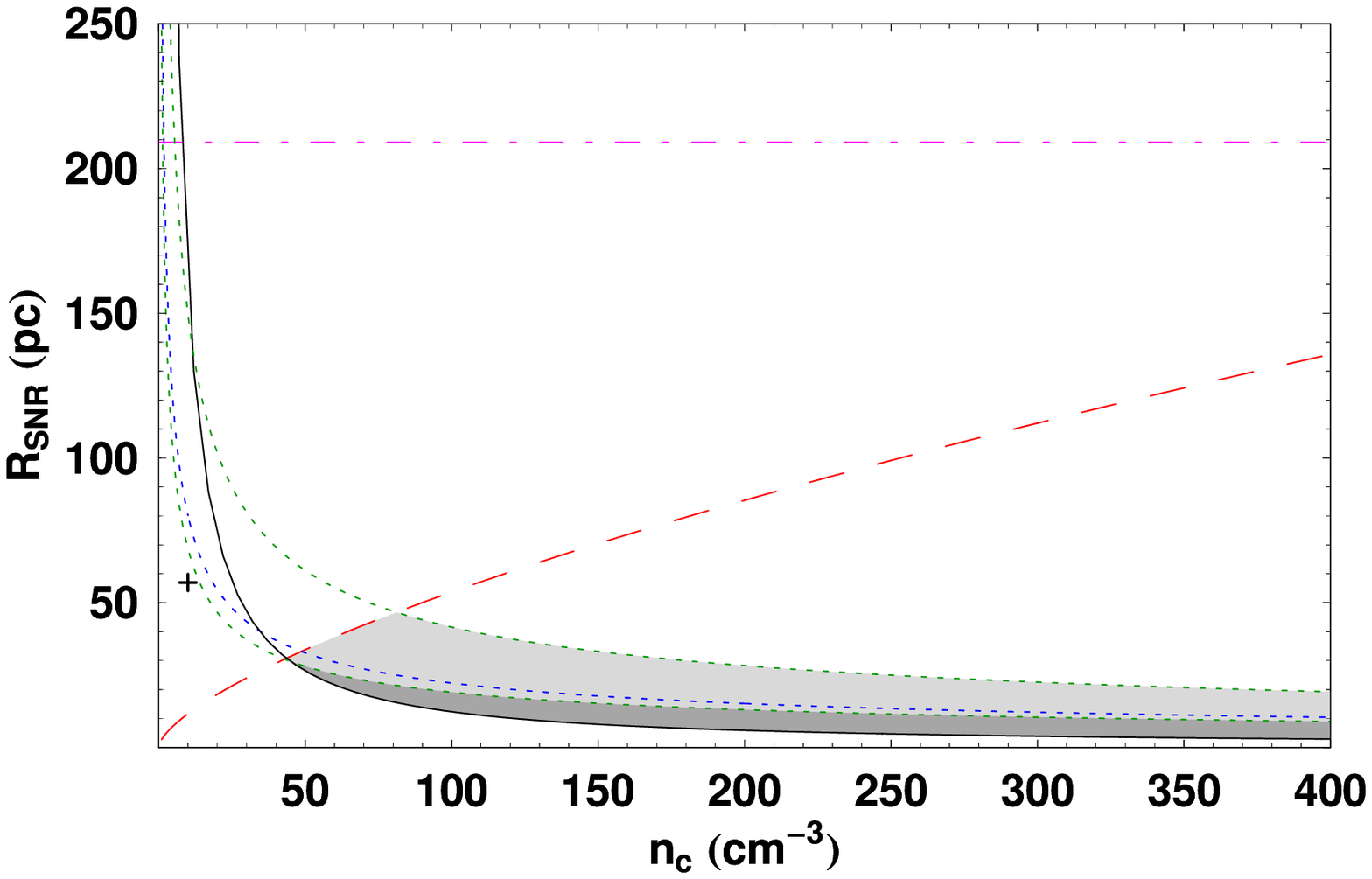}
        \includegraphics[width=6cm]{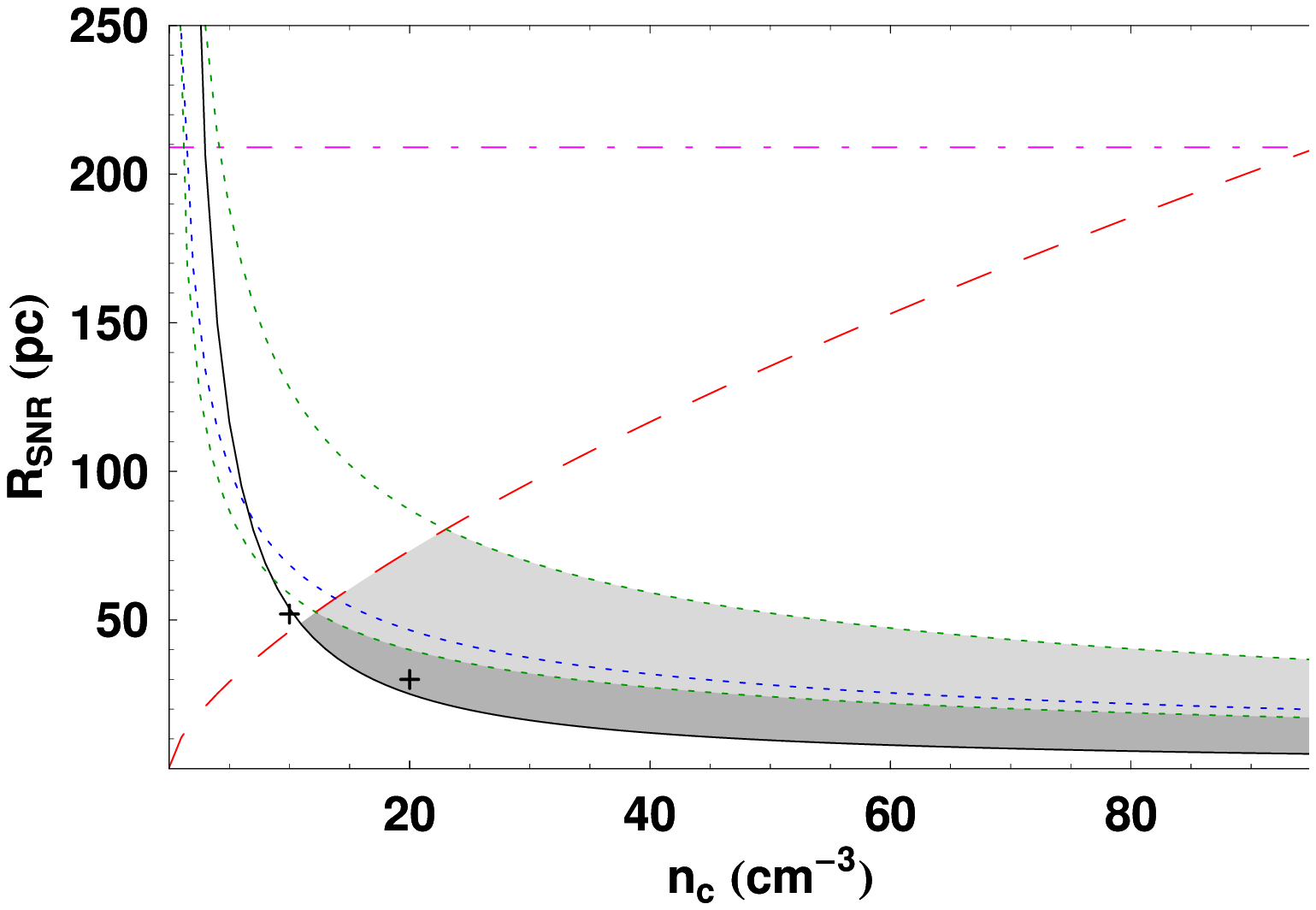}
        \includegraphics[width=6cm]{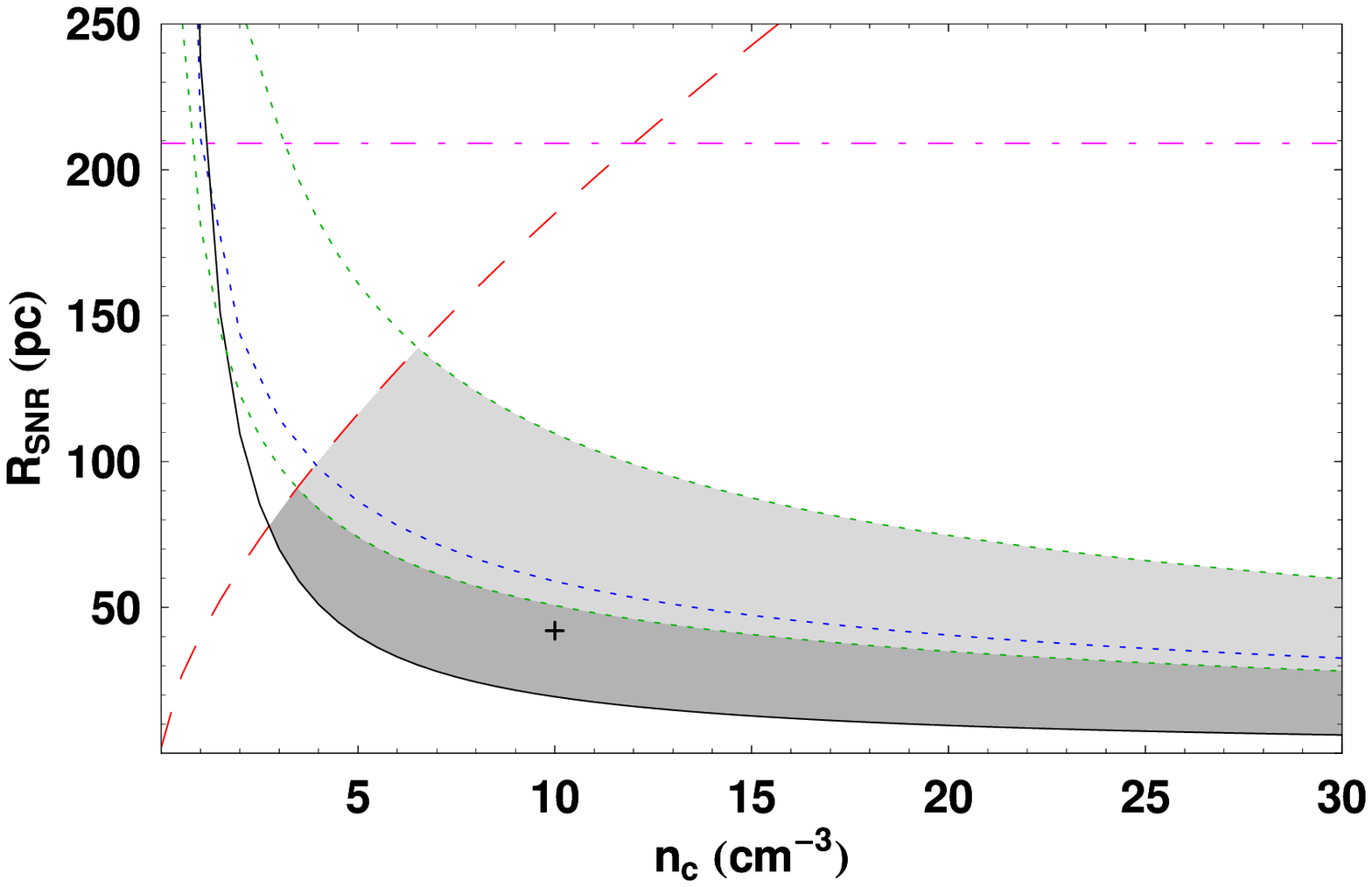}
\caption{Constraints on the SNR radius  versus cloud density for 4
different  cloud radius. Top panel: $r_c$ = 1 pc; second panel from
top: $r_c$ = 5 pc; third panel: $r_c$ = 10 pc; and bottom panel:
$r_c$ = 20 pc. Solid (black) line: upper limit for complete cloud
destruction after an encounter with an adiabatic SNR derived from
Eq. 28; dashed (red) line: upper limit for the shocked cloud to
reach the Jeans mass derived from Eq. 25 for an interaction with an
adiabatic SNR; dotted (blue) lines: upper limits for the shock front
to travel into the entire cloud before being decelerated to subsonic
velocities derived from Eqs. 30 to 34 for different values of the
cooling function $\Lambda(T)=  10^{-25}$ (lower curve), $5 \times
10^{-26}$ (middle curve), and $3 \times 10^{-27}$ erg cm$^{3}$
s$^{-1}$ (upper curve); dotted-dashed (pink) line: maximum radius
reached by a SNR in an ISM with density n=0.05 cm$^{-3}$ and
temperature T=$10^4$ K derived from Eq. 9. The shaded area defines
the region where star formation can be induced by a SNR-cloud
interaction (between the solid, dashed and dotted lines). The dark
shaded zone is bounded by the $\Lambda(T)=  10^{-25}$ erg cm$^{3}$
s$^{-1}$ dotted (blue) curve, while the light shaded zone is bounded
by the $\Lambda(T)= 3 \times 10^{-27}$ erg cm$^{3}$ s$^{-1}$ dotted
(blue) curve. The crosses in the panels indicate the initial
conditions assumed for the clouds in the numerical simulations
described in Section4.2 of the original manuscriptce}
\end{figure}

Few remarks are in order with regard to the Figure:

\begin{enumerate}
\item
In the original manuscript, the dotted (blue) curves of the diagrams
were built only for one value of the radiative cooling function of
the shocked material, i.e. $\Lambda(T) \simeq 10^{-27}$ which is
valid for a diffuse gas with a temperature $T=$ 100 K and ionization
fraction $ \le$ 10$^{-4}$. Considering that the constraint
established by the dotted (blue) curve in the diagrams is highly
sensitive to the parameter (through Eq. 31) which in turn, can vary
by two orders of magnitude depending on the value of the ionization
fraction of the cloud gas, we have presently plotted in the diagrams
three different dotted (blue) curves in order to cover a more
realistic range of possible ionization fractions from  0.1 to
10$^{-4}$, corresponding to $\Lambda(T)=  10^{-25}$ (lower dotted
curve), $5 \times 10^{-26}$ (middle curve), and  $3 \times 10^{-27}$
erg cm$^{3}$ s$^{-1}$ (upper dotted curve), respectively (see
Dalgarno \& McCray 1972). The lower dotted (blue) curve (larger
ionization fraction) bounds the dark shaded zone, while the upper
one (smaller ionization fraction) bounds the light shaded zone of
the diagrams. The middle dotted (blue) curve corresponds to the
average value of $\Lambda$ in the range above, $5 \times 10^{-26}$
erg cm$^{3}$ s$^{-1}$, and could be taken as a reference.

\item
In the solution presented in the original manuscript for the cloud
with $r_c=$ 1 pc (top panel of the Figure), there was no permitted
shaded zone where induction of star formation by SNR-cloud
interaction would be allowed. According to the present corrections
and modifications, we see that a very thin shaded "star-formation
unstable" zone appears now when the cooling function $\Lambda$ has
values which are smaller than $5 \times 10^{-26}$ erg cm$^{3}$
s$^{-1}$, or ionization fractions $< 10^{-3}$.

\item
The cross labeled  in the bottom panel of the Figure for a $r_c=$ 20
pc diffuse cloud corresponds to the initial conditions of the
numerical simulations presented in Figure 6 of the original
manuscript (that is, for a SNR at a distance $R_{snr} \sim$ 42 pc
from the surface of the cloud). In the original manuscript, that
cross lies outside the unstable shaded zone just above the upper
limit for a complete shock penetration into cloud (the dotted, blue
line of the diagram). With the present modifications, the cross now
lies near the upper limit of the dark shaded unstable zone  for
values of the cooling function $\Lambda  \lesssim   10^{-25}$ erg
cm$^{3}$ s$^{-1}$, or ionization fractions $ \lesssim 10^{-1}$. This
result remarks  how sensitive the analytical diagrams are to the
choice of $\Lambda$ (or the ionization fraction) for a given initial
temperature cloud. According to the radiative cooling
chemo-hydrodynamical simulations of Figure 6 of the original
manuscript (which corresponds to the cross in the diagram), the SNR
shock front really stalls within the cloud before being able to
cross it completely and after all the shocked cloud material does
not reach the conditions to become Jeans unstable, as predicted, but
the dense cold shell that develops may fragment and later generate
dense cores, as suggested in the original manuscript. This points to
an ambiguity of the results due to their sensitivity to $\Lambda$
and the real ionization fraction state of the gas. We should also
remark that the constraint established by the dotted (blue) curves
in the diagrams is actually only an upper limit for the condition of
penetration of the shock into the cloud. In order to estimate the
density of the shocked material at  the time $t_{st}$ when the shock
stalls within the cloud (see Eq. 31 of the original manuscript or
Eq. 15 of this Erratum), we have assumed pressure equilibrium
between the shocked  and the unshocked cloud material. A quick exam
of the numerical simulations of Figure 6, however, shows that the
shock front stalls  before this balance is attained. This implies
that the time $t_{st}$ should be smaller and therefore, the dotted
(blue) curves in the diagrams should lie below the location
predicted by Eq. (31).

\end{enumerate}

In spite of the important alterations above in the diagrams of
Figure 3, the main results and conclusions of the original
manuscript of Melioli et al. (2006), particularly those regarding
the young stellar system $\beta$ Pictoris,  remain unchanged.
Nonetheless, the present changes will be significant when compared
with similar diagrams built taking into account the effects of the
magnetic fields in the clouds. These will be presented in a
forthcoming manuscript (Le\~ao, de Gouveia Dal Pino and Melioli
2007, in prep.).

\section*{Acknowledgments}

C.M., E.M.G.D.P, and M.R.M.L. acknowledge financial support from the Brazilian Agencies FAPESP and CNPq.

\bsp

\label{lastpage}

\end{document}